# Community need for an Astrobiology Sample Repository and Sample Reference Suite

[1]A. Pontefract*, [1]C. Bradburne, [1]K. Craft, [1]M. Ballarotto, [1]K. Bowden, [1]R. Klima, [1]S. MacKenzie, [1]J. Núñez, [1]E.G. Rivera-Valentín, [2]C.E. Carr, [3]L. Chou, [4]A. Davila, [3]J.L. Eigenbrode, [5]C.R. German, [3]H.V. Graham, [6]S.S. Johnson, [4]T.M. Hoehler, [7]A.E. Murray, [3]M. Neveu, [8]B.L. Nunn, [9]H.M. Sapers, [3]S. Shkolyar, [10]A. Steele, [3]M.G. Trainer.

*Corresponding Author Email: alexandra.pontefract@jhuapl.edu. [1]Space Exploration Sector, JHU Applied Physics Laboratory, 11000 Johns Hopkins Rd., Laurel, MD, 20723. [2]School of Earth & Atmospheric Sciences, Georgia Institute of Technology, Atlanta, GA. [3] NASA/GSFC Mail Code 690/691 Greenbelt, MD 20771. [4] NASA Ames Research Center-Exobiology Branch // MS 239-4, Moffett Field, CA 94035. [5] Woods Hole Oceanographic Institution, Woods Hole, MA 02543. [6]Dept. of Biology, Georgetown University, Washington, D.C. 20015. [7]Div. Earth & Environmental Sciences, Desert Research Institute, Reno, NV 89512. [8]Dept. of Genome Sciences, University of Washington, Seattle, WA 98195. [9]Dept. of Astronomy and Planetary Sciences, Northern Arizona University, Flagstaff, AZ, 86011. [10]Carnegie Institution for Science, Washington, D.C. 20007

**Abstract:** *As we prepare for the next planetary mission charged with finding life beyond Earth, the Astrobiology community must continue to improve its understanding of what constitutes a biosignature, through the use of planetary analog samples. The study of these collected and generated samples is expanding our knowledge of what constitutes habitable environments, what life is capable of, and importantly, how to make biosignature detections within compositionally complex samples - aiding in the development of life detection instrumentation. And yet the full potential of these samples remains untapped. While the Astrobiology community possesses an incredible inventory of planetary analog samples, some incredibly precious, these are scattered across the country in individual freezers with varying degrees of documentation, curation practices, and contamination control. We, as a community, need to change the status quo of how we approach planetary analog research. One of the biggest actions we could take over the next 10 years to change that paradigm would be the creation of a sample repository for Astrobiology relevant materials, providing a centralized, well-curated, wealth of precious samples for the community. Such a collection would create a framework for material and meta-data submission that minimizes burden on the individual PIs, satisfying open data requirements; it would facilitate biosignature research, and aid in the creation of a robust life detection framework; support the development of a standardized sample reference suite for life detection instrumentation; and finally, aid in the development of techniques to be used for future sample return endeavors.*

**Introduction:** The identification and interpretation of biosignatures is a major focus in the 2019 National Academy of Sciences Astrobiology Strategy, and the 2023 National Academy of Sciences Decadal Strategy, Origins, Worlds and Life (OWL) [1, 2]. This stems from advancements in Astrobiology ranging from the detection of silicate particles and macromolecular organic material in the Enceladus plume [3, 4], to the longevity of subsurface aqueous systems on Earth, and by extension, other terrestrial planetary environments [2]. Alongside these scientific advancements, has been a corresponding increase in the development of life detection strategies and instrumentation, with a focus on agnostic detection capabilities that can work across a wide array of sample types [2]. To support these efforts, there has been a growing, recognized need in the Astrobiology community (e.g., AbSciCon 2024 Town Hall, Sample Curation [5]) for a sample repository geared toward validating life detection science and instrument engineering efforts, as well as other planetary flight investigations on environmental contextual information. In 2022, NASA outlined new Open Science requirements (i.e., SPD-41a) for PDS funded users that, in addition to data storage requirements, now additionally requires preservation of physical samples [6].

The Astrobiology community in the United States alone comprises hundreds of researchers, many of whom have conducted, or are actively conducting, terrestrial analog work across the globe in support of their investigations. These analog field sites are geographically diverse, serving a range of purposes and comprising a multitude of sample types. A significant portion of terrestrial analog research is conducted in remote and environmentally extreme locations that are both costly and logistically difficult to access, such as deep-submarine hydrothermal systems, or the Atacama, Arctic, and Antarctic – locations that provide arid, cold, and hypersaline conditions, with minimal weathering and anthropogenic contamination. Given the limitations in access to these types of sites, especially in the case of Antarctica, high fidelity samples from these locations are a precious resource for the community. In addition to terrestrial analog samples, researchers are also creating laboratory generated analog materials such as ice and regolith simulants [7], shocked lithologies [8], and synthetic biology [9], etc. The study of these collected and generated samples is expanding our knowledge of what constitutes habitable environments, what life is capable of, and importantly, how to make biosignature detections within compositionally complex samples - aiding in the development of life detection instrumentation. And yet, the full potential of these samples remains untapped. While the Astrobiology community possesses an extraordinary inventory of planetary analog samples, some incredibly precious, these are scattered across the country in individual freezers with varying degrees of documentation, curation practices, and contamination control.

***Why are these samples important to the Community?*** The current climate of research in the United States necessarily favors investigations that can conduct meaningful science under a more limited budget. It follows, then, that science that can utilize an existing sample suite (i.e., not requiring field work to collect/re-collect pristine samples), would be advantageous. Existing samples suites thus represent a wealth of potential investigations into biosignatures and related research that are not being realized, because these samples are not tracked on a community level, nor are they readily available in any meaningful way. Beyond fundamental research, the next generation of planetary missions will be focused on, and contain instrumentation dedicated to, discovering life beyond Earth. An important aspect of this type of instrument development is the ground truthing of detection capability limits, as well as potential testing across a wide range of sample types and targets, depending on the agnostic nature of the system. Having well-curated, pristine samples, available to scientists would enrich the Astrobiology community, supporting lower-cost high-impact science, furthering the state of knowledge, and supporting instrument development. A central repository would also be consistent with current NASA facilities and best practices, which curate valuable extraterrestrial sample access for science and engineering studies [10].

***Current hurdles to achieving this vision?*** With the release of SPD-41a, NASA has increased its efforts to promote open science practices which, along with data archiving, now pertains to physical sample archiving. There are, however, some major hurtles for us to address if we are to realize this opportunity: **1) A sample is only as good as its documentation**. For many investigations, knowing that you have a pristine sample, with minimal-to-no contamination is of the utmost importance. In addition, knowing what the storage conditions have been throughout its lifetime, may be a determining factor on whether or not the sample still has "value". **2) Making a sample available**. We need to know what samples are available, how much can be utilized, where they are, and then set up individual agreements with PIs (and/or memorandums of understanding) for use of their collected samples. The current practice of having each PI be responsible for maintaining their own samples, while cost-saving from an immediate standpoint, is ultimately only a short-term solution that is nonetheless highly problematic. Firstly, a documented sample-chain is an important part of any laboratory's standard operating procedure: starting at the conditions of collection, through transportation, and finally to the permanent storage location. Unfortunately, unless the samples are USDA (United States Department of Agriculture) controlled – requiring government mandated detailed record keeping and well-maintained laboratory facilities for the use of the collected sample – the final storage location is frequently where sample tracking ends. Undocumented sub-sampling, temperature cycling, and chain of custody changes for example, may occur during the lifetime of a sample in permanent storage, all without any documentation. This is not to imply that laboratories are not well-intentioned in their sample curation, more that the current intent of sample collection is not that of future distribution and archiving; these samples are only intended for the PI Laboratory. Secondly, the reality of an academic setting means that samples can pass through multiple iterations of (student) investigators, with varying adherence to best practices with sample handling. Finally, sample preservation is greatly affected when PIs change laboratories, and especially when they retire, likely resulting in the loss and/or destruction of samples.

*What do we need?* We, as a community, need to change the status quo of *how* we approach planetary analog research. One of the biggest actions we could take over the next 10 years to change that paradigm would be the creation of a centralized "**Open Sample Repository**" for Astrobiology relevant materials. Centralized repositories are not a groundbreaking idea, but their implementation for Astrobiology has profound and powerful implications for our field. Our inspiration comes from facilities such as the JSC Apollo Collection [11], and the NSF Ice Core [12] and NSF Sediment Core collections at Florida State [13] and Lamont Doherty [14]. These sample repositories provide a centralized, well-curated, wealth of precious samples for the community, enabling the continuation of scientific investigation and further justifying the cost expended in acquiring and storing these unique sample sets. If we decide that the lifespan of our terrestrial analog samples should outstrip the lifespan of the project it was collected for, what sort of returns on science will we then be able to realize? The 2019 NAS Astrobiology Strategy for the Search for Life identified an urgent need to develop standardized methods for assessing the predictive value of biosignatures, and that the path forward was "*to strive for a comprehensive, quantitative foundation that uses multiple lines of evidence and environmental context to provide the most robust life detection framework involving interdisciplinary laboratory, field, and modeling work, as well as community efforts to develop a consensus on assessment to apply to the Search for life Beyond Earth."*[1]. A centralized sample repository would provide a path forward to reaching that assessment. Critically, a centralized sample repository could also create a framework for material and meta-data submission that minimizes burden on the individual PIs.

*Sample curation for the Astrobiology community:* We propose the funding of a centralized facility (or set of facilities) that will serve as a repository for PSD funded users to send a subset of their collected/generated planetary analog samples to then be made available to the community (**Fig. 1**). Precious and unique samples could then serve a renewed purpose, while simultaneously satisfying the requirements of SPD-41a. It is plausible that a single lone facility may not be the best approach due to the potential size required for such an installation, as well as significant differences in storage requirements across the breadth of potential samples; however, a centralized set of locations where sample ingress, storage, tracking, and egress are a highly controlled process, following NPR 7100.5 [15] and ISO 17025 standards for testing and calibration laboratories, would be critical to ensure success. These samples would be available for viewing on a centralized database, along with their accompanying metadata. Users would then be able to mine curated metadata, and can also request physical samples for investigation on a case-by-case basis. Sample governance would cover issues of permitting, patent filing, authorship, embargoes, sponsor open data requirements and/or links. Lastly, an initial effort would define a set of curation practices and data libraries that would allow broad interconnection with other sample repositories, curating extraterrestrial materials [11], terrestrial microbiology [16], geologic specimens [17], and others as defined by the community. This could allow interconnection and metadata and sample comparisons to occur with existing repositories around the nation and potentially the world.

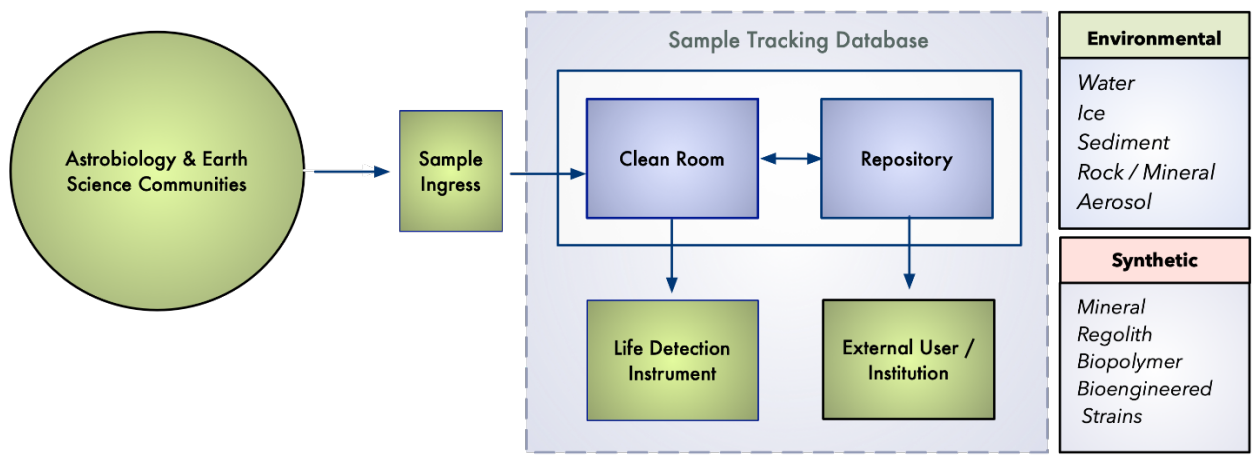

*Figure 1. Concept facility for the Astrobiology and Planetary Analogs Sample Collection. Arrows detail sample ingress and egress, as well as the role of the clean room in sample processing as well as instrument testing. Examples of potential sample types are listed on the right, spanning environmental analog samples to synthetic (lab-generated) samples.*

***Beyond curation: a sample reference suite for life detection***. There have been long-standing community-led conversations surrounding the utility and necessity of a standardized sample reference suite for life-detection instrumentation in order to provide a baseline against which to evaluate performance and validate capability (e.g., NFoLD). These conversations have previously stagnated due to a lack of direct funding to support such an effort, as well as disagreements on what such a reference set would look like: e.g., is it possible to capture the breadth of possible sample types? How do you handle sample heterogeneity? Forming an Astrobiology sample repository would be a huge step to realizing this goal, where the community would have a curated set of samples from which we could begin to down-select to form such a reference suite on a case-by-case basis. Moreover, such a facility would be well situated to store such a special collection, and could even be designed to allow for on-site testing of life-detection instrumentation under clean-room conditions (e.g., ISO 5-7).

*Next steps?* Beyond likely the biggest question of funding, there are many other questions that need to be addressed in moving forward with this concept: What is the lifetime size of the proposed collection? Which samples are the highest priority? Is there a minimum sample volume that would be required to justify curation? Who presides over the distribution of samples? How do federal lands (e.g., national park, and traditional lands) 'ownership' affect sample curation and availability? Etc. The community would highly benefit from a series of workshops geared towards addressing these questions, so that we may collectively agree upon and be part of, this ambitious endeavor. A mandate of the 2023 Decadal Strategy is to understand *"what sorts of signatures indicate life (or the absences of life) and how to avoid both false positives and false negatives"* [2]. A centralized sample repository with standardized curation practices for analog samples would facilitate those investigations by preserving the hard work of collecting or generating the materials for generations to come. It could also serve as a resource to enable science and engineering development for the study and discovery of life beyond Earth, enabled by the community availability of standardized, high-fidelity astrobiological analogues. As we look toward coming decades, we must also think about the skills that will be required to handle returned samples, such as with Mars Sample Return; such a collection could allow us to develop and hone the strategies and techniques we will need in the study of such priceless samples.